# Anatomy of Dzyaloshinskii-Moriya Interaction at Co/Pt Interfaces

Hongxin Yang[1,2], André Thiaville[2], Stanislas Rohart[2], Albert Fert[3], & Mairbek Chshiev[1*]

[1] *Univ. Grenoble Alpes, INAC-SPINTEC, F-38000 Grenoble, France; CNRS, SPINTEC, F-38000 Grenoble, France; and CEA, INAC-SPINTEC, F-38000 Grenoble, France*

[2] *Laboratoire de Physique des Solides, Université Paris-Sud, CNRS UMR 8502, F-91405 Orsay Cedex, France*

[3] *Unité Mixte de Physique CNRS/Thales, 1 Av. Fresnel, 91767 Palaiseau, France and Université Paris-Sud, 91405 Orsay, France*

*e-mail: mair.chshiev@cea.fr

**The Dzyaloshinskii-Moriya Interaction (DMI) between spins is induced by spin-orbit coupling in magnetic materials lacking inversion symmetry[1-3]. DMI is recognized to play a crucial role at the interface between ferromagnetic (FM) and heavy nonmagnetic (NM) metals to create topological textures called magnetic skyrmions[4-6] which are very attractive for ultra-dense information storage and spintronic devices[7-9]. DMI also plays an essential role for fast domain wall (DW) dynamics driven by spin-orbit torques[10-12]. Here, we present first principles calculations which clarify the main features and microscopic mechanisms of DMI in Co/Pt bilayers. DMI is found to be predominantly located at the interfacial Co layer, originating from spin-orbit energy provided by the adjacent NM layer. Furthermore, no direct correlation is found between DMI and proximity induced magnetism in Pt. These results clarify underlying mechanisms of DMI at FM/NM bilayers and should help optimizing material combinations for skyrmion- and DW-based storage and memory devices.**

The discovery of fast current-controlled domain wall motion induced by spin-orbit torques in perpendicularly magnetized thin ferromagnetic layers deposited on NM metals of large spin-orbit (SO) coupling[13] is promising for the development of novel memory and storage devices with high density, performance and endurance[14]. The origin of the current-controlled DW motion was first attributed to Rashba effect[13,15-17] and later to Spin Hall effect (SHE)[18,19]. In 2012 it was shown by micromagnetic calculations[10], and later confirmed experimentally[11,12], that the role of Dzyaloshinskii-Moriya interaction (DMI)[1-3] at the FM/NM interface was essential to stabilize the DWs in a Néel configuration with a given chirality, thus allowing their fast motion by SHE in a direction fixed by the chirality. The DMI is also involved in creating and stabilizing magnetic skyrmions in magnetic thin films deposited on heavy metals[6]. A calculation and understanding of interface DMIs is therefore important to elucidate its role in magnetization dynamics or generation of skyrmions, and to assess the potential of the recently proposed DW- or skyrmion-based memory and storage concepts[7-9,14].

In this Article we use *ab initio* calculations on Pt/Co bilayers to evaluate both the total DMI of the bilayer and its distribution in the successive atomic layers of Co and Pt (Methods). The DMI energy restricted to nearest neighbour normalized spins (denoted $\mathbf{S}_i$ and $\mathbf{S}_j$) can be written as

$$E_{DMI} = \sum_{<i,j>} \mathbf{d}_{ij} \cdot (\mathbf{S}_i \times \mathbf{S}_j) \qquad (1)$$

with summation over bonds involving DMI vectors $\mathbf{d}_{ij}$ for two types of pairs, those inside a given layer $k$, and interlayer pairs between a layer $k$ and layers above/below. It turns out that the contributions from interlayer pairs are very small (Supplementary S1), so that, in first approximation, we will neglect them. From the Moriya symmetry rules[2], the DMI vector for the layer $k$ can be written as $\mathbf{d}_{ij}^k = d^k(\hat{\mathbf{z}} \times \hat{\mathbf{u}}_{ij})$ where $\hat{\mathbf{z}}$ and $\hat{\mathbf{u}}_{ij}$ are unit vectors pointing along z and from $\mathbf{S}_i$ to $\mathbf{S}_j$, respectively. The total DMI strength, $d^{tot} = \sum_k d^k$, is derived (Supplementary S1) by identifying the difference between the DFT energies $E_{CW}$ and $E_{ACW}$ for opposite chirality spin configurations as shown in Fig. 1(a) and (b) with the corresponding energy differences calculated from equation (1):

$$d^{tot} = (E_{CW} - E_{ACW})/12 \qquad (2)$$

The $d^{tot}$ can be seen as the DMI strength concentrated in a single atomic layer and producing an equivalent effect (at least if the total thickness is smaller than the exchange stiffness length). The global effect on the bilayer can also be expressed by the micromagnetic energy per unit volume[7-10]:

$$E = D[m_z \frac{dm_x}{dx} - m_x \frac{dm_z}{dx}] + id \cdot (x \leftrightarrow y) \tag{3}$$

where a coefficient $D$ is related to $d^{tot}$ as (Supplementary S2 and S3):

$$D = \frac{3\sqrt{2} d^{tot}}{N_F a^2} \tag{4}$$

where $a$ is the fcc lattice constant.

For the total DMI strength $d^{tot}$ of hcp(0001)Co/fcc(111)Pt bilayers we find a large value in the range 1.5-3 meV, of anticlockwise chirality, with a dependence on the thickness of Co and Pt shown in Fig. 2(a). The dependence on Co thickness, as explained by layer-resolved results shown below, comes from the small but not negligible extension of the DMI to other Co layers away from the interface Co layer. Except for the Pt(n)Co(1) series, the influence of the Pt thickness is weak. Globally $d^{tot}$ tends to an approximately constant value at large thickness which is consistent with the interface character of the DMI. $D$, as it corresponds to an average of the DMI in the Co film, shows the expected decrease with the Co thickness. Finally, we have also studied the effect of intermixing between Co and Pt. When one Co atom is swapped with Pt at interface (25% interfacial mixing), the total DMI is decreased by half (open and blue circles on Fig.2).

In order to clarify the physics of the interface DMI, we calculated the layer-resolved DMI vector amplitude $d^k$ by considering spin configurations as those of Fig. 1(c) and (d) with opposite chirality in a single layer while spins in all other layers are constrained to be along $y$ (Supplementary S1.A). The corresponding DFT energies $E^k_{CW}$ and $E^k_{ACW}$ allow finding $d^k$ from Eq. (2). The results for hcp(0001)Co(3ML) on fcc(111)Pt(3ML) are shown in Fig. 3(a). As a test of the accuracy of our approach, we have checked that the sum of the $d^k$ of the different layers is close to $d^{tot}$ (Fig.2a) with the slight difference (~10% of $d^{tot}$) caused by the aforementioned interlayer contributions integrated into $d^{tot}$ but not into $d^k$ (Supplementary S1), and by the fact that DFT calculations on differently constrained configurations cannot be strictly equivalent. We will show now that the physics of DMI is cleared up by the distribution of the $d^k$ and the corresponding distribution of electronic energy local differences between configurations of opposite chiralities, $\Delta E_{SOC}$ (Fig. 3).

The most obvious feature of the distribution of $d^k$ in Fig. 3(a) is that the DMI is predominantly located at interfacial Co layer, as indicated by the blue bar on Co1, with definitely smaller and opposite DMIs in Co2 and Co3, and a much smaller contribution in Pt1 (almost nothing in Pt2 and Pt3). It is interesting to see where is located the difference $\Delta E_{SOC}$ in spin-orbit coupling (SOC) associated with opposite chiralities in the Co1 layer (see Supplementary S4). As shown in Fig. 3(b), the large DMI at Co1 interface layer is associated with a large electronic energy change $\Delta E_{SOC}$ in the adjacent Pt layer, consistently with the Fert-Levy picture[3]. Considering the Co3 layer, as Co3 is too far from Pt, its small DMI is associated to (small) SOC energy changes in Co2 and Co1. Similarly, the DMI of Co2 takes its origin from both adjacent Co layers and the moderately distant Pt1 and Pt2 layers (red bars). The DMI is very small for the proximity-induced spins (about $0.3\mu_B$) in the interfacial Pt1 monolayer with a SOC energy originating mainly from other Pt layers (magenta bars). The SOC energy distribution associated with the total DMI $d^{tot}$, shown in inset of Fig. 3(b), is very similar to that obtained for $d^{Co1}$. Some features of the total DMI thickness dependence can also be understood. For instance, one sees that the small DMI of Co2 and Co3 is opposite to the large DMI of Co1, which explains the smaller $d^{tot}$ in Co2Pt3 and Co3Pt3 compared to Co1Pt3 in Fig. 2(a).

It was suggested[20] that DMI at FM/NM interfaces is directly related to the existence of a proximity induced magnetic moment (PIM) in NM. To test this suggestion, we derived the DMI as a function of PIM by constraining the magnitude of the latter in the Pt1 layer of a Pt3/Co3 structure. Actually, as shown in Fig.4a, we find the opposite result: a Pt moment reduction increases the DMI, as well as $d^{tot}$ and $d^{Co1}$. The DMI is maximum at about zero Pt moment with a decrease of 31% for $d^{tot}$ when PMI increases from zero to 0.4 $\mu_B$. Consistently, in spite of a larger moment on Pd in Pd/Co compared to Pt in Pt/Co, we find a definitely smaller DMI for Pd/Co (Fig.2), which indicates that the essential factor is the SOC (larger in the Pt 5$d$ states than in the Pd 4$d$ states).

For Co/Pt(111), our DMI values are in agreement with recent theoretical (1.75 meV for Co3/Pt3(111)[21]) and experimental[22] reports, but distinctly smaller than the DMI calculated in a Berry phase approach[23]. As a comparison study, we find that, in Co/Au, DMI is much weaker and of opposite chirality compared to that in Co/Pt (cf. Fig. 2). The origin of this difference of DMI between these similar systems can be attributed to the absence of strongly spin-orbit coupled $d$ states at the Fermi level in Au yielding a strong reduction of $\Delta E_{SOC}$ on Au1 (Fig. 4(b)). We have also calculated DMI at Fe/Ir and Co/Ir interfaces. We find a very large clockwise DMI (-1.9 meV) for a single layer of Fe on 3 layers of Ir(111), in very good agreement with Ref. 24. The DMI we find for Co/Ir (-0.22 meV) is also of clockwise chirality but smaller than Fe/Ir (Fig. 2). The difference between Co/Ir and Fe/Ir cannot be explained by simple arguments and requires further analysis. The finding of opposite chiralities for Pt and Ir when both are below or both above Co leads to the interesting prediction of additive effects and large DMI when Co is between Pt and Ir.

To sum up, we have used first principles calculations to determine the DMI in Co/Pt bilayers and cleared up its physical mechanism. Our main conclusion is that the large anticlockwise DMI of the bilayers (~3 mJ/m$^2$ for Pt3/Co3) has a predominant contribution from pair couplings between the spins of the interfacial Co layer. This DMI between the interface Co spins is directly related to the change of the SOC energy in the interface Pt atoms when the Co spin chirality is reversed. The DMI does not extend significantly into other Co layers and is very weak between the proximity-induced spins in Pt. We have also shown that the DMI of the Co/Pt bilayers is not related to the existence of proximity-induced magnetism in Pt. Our similar calculations of DMI for the Fe/Ir system are in agreement with previous ab-initio calculations. The smaller DMIs we find for Co/Pd and Co/Au can be respectively explained by the smaller SOC of the $d$ states in Pd and the absence of $d$ states at the Fermi level in Au.

**METHODS**

The Vienna ab initio simulation package (VASP) was used in our calculations with electron-core interactions described by the projector augmented wave method for the pseudopotentials, and the exchange correlation energy calculated within the generalized gradient approximation of the Perdew-Burke-Ernzerhof (PBE) form[24,25]. The cutoff energies for the plane wave basis set used to expand the Kohn-Sham orbitals were chosen to be 320 eV for all the calculations. The Monckhorst-Pack scheme was used for the $\Gamma$-centred $4\times16\times1$ k-point sampling. In order to extract the DMI vector, calculations were performed in three steps. First, structural relaxations were performed until the forces become smaller than 0.001 eV/Å for determining the most stable interfacial geometries. In our DMI calculations, we use 1 to 3 layers of Co on 1 to 3 layers of Pt(Au) film in a 4 by 1 surface unit cell with $\pi/2$ spin rotations (Fig. 1). Of note, one could use a unit cell with different spin spiral period without affecting the method validity, as explained in Supplementary S3. Next, the Kohn-Sham equations were solved, with no spin-orbit coupling (SOC), to find out the charge distribution of the system's ground state. Finally, SOC was included and the self-consistent total energy of the system was determined as a function of the orientation of the magnetic moments which were controlled by using the constrained method implemented in VASP.

This method has been used for DMI calculations in bulk frustrated systems and insulating chiral-lattice magnets[26,27], and was adapted here to the case of interfaces. Details of the model are described in Supplementary material.

**REFERENCES**

1. Dzialoshinskii, I. E. Thermodynamic theory of 'weak' ferromagnetism in antiferromagnetic substances. *Sov. Phys. JETP* **5**, 1259–1262 (1957).

2. Moriya, T. Anisotropic superexchange interaction and weak ferromagnetism. *Phys. Rev.* **120**, 91–98 (1960).

3. Fert, A. & Levy, P. M. Role of anisotropic exchange interactions in determining the properties of spin glasses. *Phys. Rev. Lett*. **44**, 1538–1541 (1980).

4. Mühlbauer, S. *et al*. Skyrmion Lattice in a Chiral Magnet, *Science* **323,** 915–919 (2009).

5. Yu, X. Z. *et al*. Real-space observation of a two-dimensional skyrmion crystal, *Nature* **465,** 901–904 (2010).

6. Heinze, S. *et al*. Spontaneous atomic-scale magnetic skyrmion lattice in two dimensions, *Nature Phys.* **7**, 713–718 (2011).

7. Fert, A., Cros,V. & Sampaio, J. Skyrmions on the track, *Nature Nanotech*. **8**, 152–156 (2013).

8. Sampaio, J., Cros, V., Rohart, S., Thiaville, A. & Fert, A. Nucleation, stability and current-induced motion of isolated magnetic skyrmions in nanostructures. *Nature Nanotech*. **8**, 839-844 (2013).

9. Iwasaki, J., Mochizuki, M. & Nagaosa, N. Current-induced skyrmion dynamics in constricted geometries. *Nature Nanotech*. **8**, 742-747 (2013).

10. Thiaville, A., Rohart, S., Jué, E., Cros, V. & Fert, A. Dynamics of Dzyaloshinskii domain walls in ultrathin magnetic films. *Europhys. Lett*. **100**, 57002 (2012).

11. Emori, S., Bauer, U., Ahn, S.-M., Martinez, E. & Beach, G. S. D. Current-driven dynamics of chiral ferromagnetic domain walls. *Nature Mater*. **12**, 611–616 (2013).

12. Ryu, K.-S., Thomas, L., Yang, S.-H. & Parkin, S. Chiral spin torque at magnetic domain walls. *Nature Nanotech.* **8**, 527–533 (2013).

13. Miron, I. M. *et al.* Fast current-induced domain-wall motion controlled by the Rashba effect. *Nat. Mater.* **10**, 419–423 (2011).

14. Parkin, S. S. P., Hayashi, M. & Thomas, L. *Science* **320,** 197202 (2009).

15. Kim, K.-W., Seo, S.-M., Ryu, J., Lee, K.-J. & Lee, H.-W. Magnetization dynamics induced by in-plane currents in ultrathin magnetic nanostructures with Rashba spin-orbit coupling. *Phys. Rev. B* **85**, 180404 (2012).

16. Miron, I. M. *et al.* Perpendicular switching of a single ferromagnetic layer induced by in-plane current injection. *Nature* **476**, 189–193 (2011).

17. Garello, K. *et al.* Symmetry and magnitude of spin-orbit torques in ferromagnetic heterostructures. *Nature Nanotech.* **8**, 587–593 (2013).

18. Liu, L. *et al.* Spin-torque switching with the giant spin Hall effect of tantalum. *Science* **336**, 555–558 (2012).

19. Hirsch, J. E. Spin Hall effect. *Phys. Rev. Lett.* **83**, 1834 (1999).

20. Ryu, K.-S., Yang, S.-H.,Thomas, L. & Parkin, S. S. P. Chiral spin torque arising from proximity-induced magnetization. *Nature Commun.* **5**, 3910 (2014).

21. Dupé, B., Hoffmann, M., Paillard, C. & Heinze, S. Tailoring magnetic skyrmions in ultra-thin transition metal films, *Nature Commun.* **5**, 4030 (2014).

22. Hrabec, A. *et al*. Measuring and tailoring the Dzyaloshinskii-Moriya interaction in perpendicularly magnetized thin films, *Phys. Rev. B* **90**, 020402(R) (2014).

23. Freimuth, F., Blügel, S. & Mokrousov, Y. Berry phase theory of Dzyaloshinskii-Moriya interaction and spin–orbit torques. *J. Phys: Condens. Matter* **26**, 104202 (2014).

24. Kresse, G. & Hafner, J. Ab initio molecular dynamics for liquid metals. *Phys. Rev. B* **47**,

558–561 (1993).

25. Kresse, G. & Furthmüller, J. Efficient iterative schemes for ab initio total-energy calculations using a plane-wave basis set. *Phys. Rev. B* **54**, 11169–11186 (1996).

26. Xiang, H. J., Kan, E. J., Wei, S.-H., Whangbo, M.-H. & Gong, X. G. Predicting the spin-lattice order of frustrated systems from first principles. *Phys. Rev. B* **84**, 224429 (2011).

27. Yang, J. H. *et al.* Strong Dzyaloshinskii-Moriya Interaction and Origin of Ferroelectricity in $Cu_2OSeO_3$. *Phys. Rev. Lett.* **109**, 107203 (2012).

**ACKNOWLEDGEMENTS**

The authors thank P. M. Levy, I. M. Miron, O. Boulle, L. Buda-Prejbeanu and G. Gaudin for fruitful discussions. This work was supported by ANR SOSPIN and ESPERADO, by a CNRS postdoctoral fellowship, and used HPC resources from CEA/Grenoble and GENCI-CINES (Grants 2012&2013-096971).

**AUTHOR CONTRIBUTIONS**

H. Y. and M. C. developed the approach and performed the calculations. M.C. prepared the manuscript with help of H. Y. and A. F. All authors discussed the results and commented on the manuscript.

**COMPETING INTERESTS STATEMENT**

The authors declare that they have no competing financial interests.

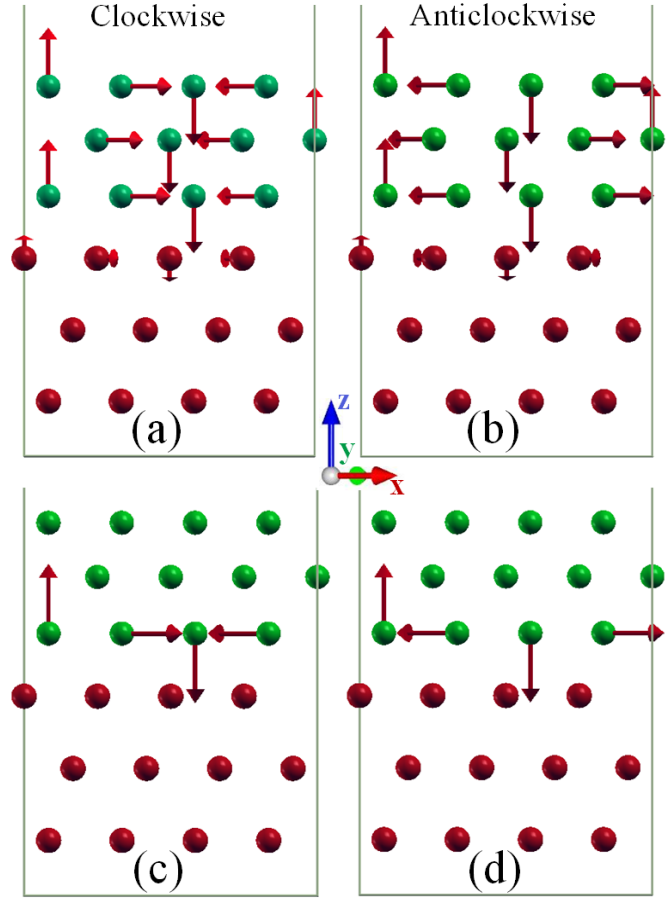

**Figure 1 Spin configurations used to calculate the DMI of hcp(0001)Co/fcc(111)Pt bilayers. a,** Clockwise (CW) and **b,** anticlockwise (ACW) spin configurations used to calculate the total DMI of the system. **c,** CW and **d,** ACW configurations with spiral configuration in a single layer and spins constrained along *y* axis used to find the layer resolved DMI parameter $d^k$. Green and red correspond respectively to Co and Pt. The spin moments of Pt atoms are multiplied by 10 for convenience. The side view of the unit cell for the Co3Pt3 case is represented here.

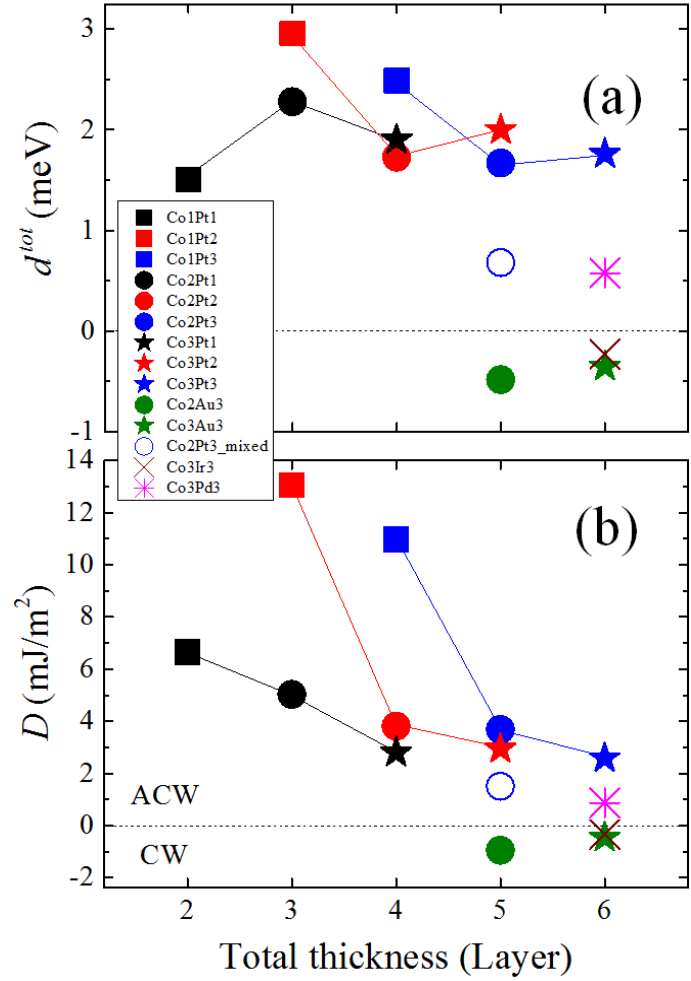

**Figure 2 DMI of Co/Pt bilayers, Co/Pt bilayer with mixed interfaces, Co/Ir, Co/Au and Co/Pd bilayers as a function of the total number of layers.** Solid lines can be used to follow the variation the DMI of Co/Pt bilayers as a function of the Co thickness for a given Pt thickness. **a,** Total DMI coefficient $d^{tot}$ and **b,** Micromagnetic DMI coefficient $D$, as a function of slab thickness. CW(ACW) indicates clockwise(anticlockwise) chirality.

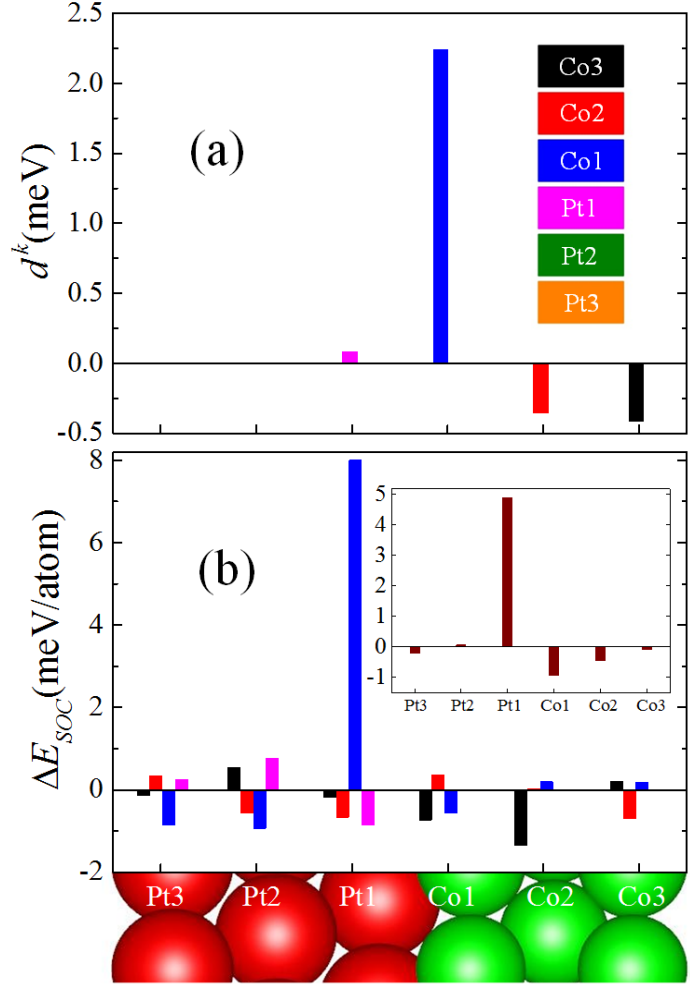

**Figure 3  Anatomy of DMI for 3 layers of Co on 3 layers of Pt.  a,** Layer resolved DMI coefficient $d^k$ of layer $k$. **b,** The corresponding localization of the associated SOC energy in the atomic sites. For example, it can be seen that the large DMI coefficient of the Co1 layer (blue bar in **a**), is associated with a large variation $\Delta E_{SOC}$ of SOC energy in the Pt1 layer (see corresponding blue bar in **b**) induced by an inversion of the chirality in Co1. For comparison the distribution of SOC energy variations induced by an inversion of the chirality of the total structure is also shown (inset).

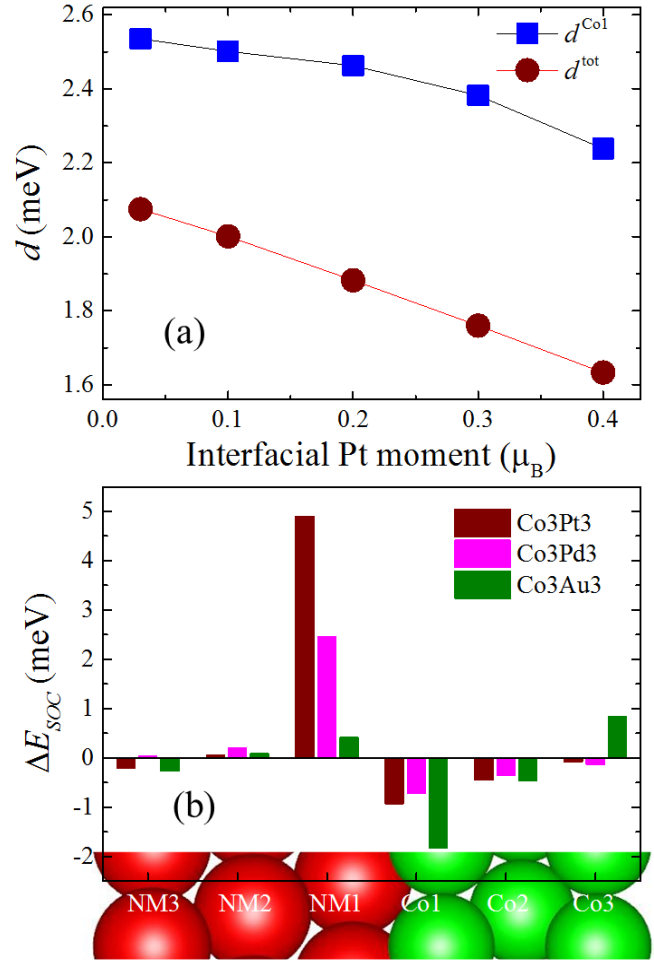

**Figure 4 Influence of the proximity induced moment on DMI for Co/Pt bilayers and localization of $\Delta E_{SOC}$ for Co/Pt, Co/Pd and Co/Au. a,** DMI parameters $d^{tot}$ (circles) and $d^{Co1}$ (squares) for Co3/Pt3 bilayers as a function the interfacial spin moment in Pt1 layer. **b,** Distribution of spin-orbit energy associated with $d^{tot}$ calculations across Co/Pt, Co/Pd and Co/Au structures. One can see that $\Delta E_{SOC}$ on NM1 sites is strongly reduced in case of Co/Pd and Co/Au compared to Co/Pt bilayers, resulting in strong reduction of DMI.